\documentclass[sn-mathphys-num,iicol]{sn-jnl}

\usepackage{graphicx}%
\usepackage{multirow}%
\usepackage{amsmath,amssymb,amsfonts}%
\usepackage{amsthm}%
\usepackage{mathrsfs}%
\usepackage[title]{appendix}%
\usepackage{xcolor}
\usepackage{textcomp}%
\usepackage{manyfoot}%
\usepackage{booktabs}%
\usepackage{algorithm}%
\usepackage{algorithmicx}%
\usepackage{algpseudocode}%
\usepackage{listings}%
\usepackage{makecell}
\usepackage{lineno}

\definecolor{Functions}{rgb}{0.176, 0.608, 0.941}
\definecolor{Gensteps}{rgb}{0.949, 0.278, 0.149} 
\definecolor{Variables}{rgb}{0.980, 0.780, 0.067} 
\definecolor{OpticksClasses}{rgb}{0.505, 0.56625, 0.161875}

\theoremstyle{thmstyleone}%
\theoremstyle{thmstyletwo}%

\theoremstyle{thmstylethree}%

\begin{document}

\title{ Performance of an Optical TPC Geant4 Simulation with Opticks GPU-Accelerated Photon Propagation }

\author[3]{I.~Parmaksiz,}\email{ilker.parmaksiz@mavs.uta.edu}
\author[3]{K.~Mistry,}
\author[21]{E.~Church,}
\author[2]{C.~Adams,}
\author[3]{J.~Asaadi,}
\author[19]{J.~Baeza-Rubio,}
\author[2]{K.~Bailey,}
\author[3]{N.~Byrnes,}
\author[3]{B.J.P.~Jones,}
\author[12]{I.A.~Moya,}
% ...
\affil[2]{
Argonne National Laboratory, Argonne, IL 60439, USA}
\affil[3]{
Department of Physics, University of Texas at Arlington, Arlington, TX 76019, USA}
% ...
\affil[12]{
Case Western Reserve University, Cleveland, OH 44106, USA}
% ...
\affil[19]{
Wright Laboratory, Department of Physics, Yale University, New Haven, CT 06520, USA}
% ...
\affil[21]{
Pacific Northwest National Laboratory (PNNL), Richland, WA 99352, USA}
% ...


\abstract{We investigate the performance of \texttt{Opticks}, a \texttt{NVIDIA OptiX API} 7.5 GPU-accelerated photon propagation tool compared with a single-threaded \texttt{Geant4} simulation. We compare the simulations using an improved model of the \texttt{NEXT-CRAB-0} gaseous time projection chamber. Performance results suggest that \texttt{Opticks} improves simulation speeds by between $58.47\pm{0.02}$ and $181.39\pm{0.28}$ times relative to a CPU-only \texttt{Geant4} simulation and these results vary between different types of GPU and CPU. A detailed comparison shows that the number of detected photons, along with their times and wavelengths, are in good agreement between \texttt{Opticks} and \texttt{Geant4}.}


\maketitle


\begin{figure*}[t!]
\centering
\includegraphics[width=0.9\textwidth]{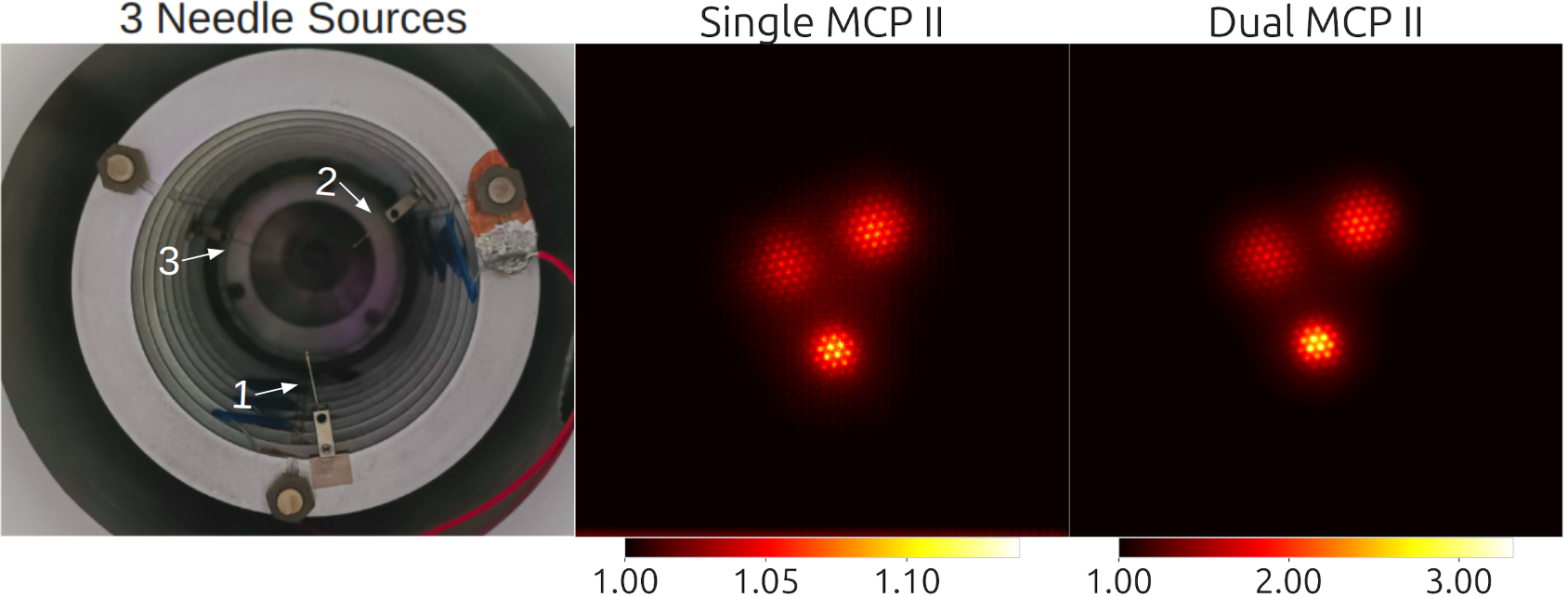}
\caption{Left: the interior of the \texttt{NEXT-CRAB-0} detector. Three needle sources are shown by the thin metal rods attached to the field cage with distances 4, 10, and 14 cm from the electroluminescence region for needles 1, 2, and 3, respectively. Center and Right: Images from the \texttt{NEXT-CRAB-0} detector with Single (center) and Dual MCP II (right) at $6$~bar xenon with three Pb-210 alpha sources. The images contain the averaged intensities from alpha events. The color of the images is normalized to their mean. }
\label{fig:Sources}
\end{figure*}

\section{Introduction}
\label{sec:intro}

Time projection chambers (TPCs) generate tens to hundreds of millions of photons through primary and secondary scintillation light when charged particles in the keV to MeV energy range interact in the detector volume. The simulation of these photons, known as \textit{optical photons}, is commonly carried out using the software package \texttt{Geant4} \cite{Allison2006,Allison2016,Agostinelli2003}. In these simulations, the photon propagation is often the largest portion of the simulation time compared with the primary particle generation, ionization electron drift, detector and physics initialization, and file writing. This is predominately due to the huge number of photons needed to be tracked through the detector geometry.

Many methods are employed to reduce simulation times with some approximations. The optical library method involves generating precomputed light tables by simulating hundreds of millions of optical photons in voxelized regions of the detector. The probability of producing a photoelectron on a sensor such as a photomultiplier tube (PMT) can then be mapped for a given energy deposition in the detector volume from a particular voxel. While this method is practical, simulation times scale with the detector volume, and it can take several weeks to produce the tables for a detector geometry. In addition, for large detectors and/or those with large numbers of sensors, 
files can grow to huge sizes and the larger memory requirements make the use of light tables unfeasible. Other methods include the semi-analytical model~\cite{Garcia-Gamez2021}, generative adversarial neural networks (GAN)~\cite{Li2023GANEXO200, Fu2024KamLANDZen, GNN_photon_Propagation} employed by multiple experiments including nEXO, KamLAND-Zen, SBND~\cite{SBND2024} and DUNE~\cite{dunecomputingtdr}, and differential simulations with ML~\cite{siren}. The semi-analytical model predicts (with a precision better than 10\%) the amount of observed light at a photosensor using the relative positions of the energy deposition and photosensor while the GAN is trained to make predictions of the number of photons at each photodetector with 20-50 times improvement with respect to \texttt{Geant4}. Differential simulations provide flexibility with tuneable parameters from learning directly on the data to help tune the agreement between data and simulation. 

In this paper, we employ the use of graphical processing units (GPUs) to propagate optical photons within a \texttt{Geant4} simulation. GPUs offer large parallelization, which can rapidly speed up simulation times, accounting for many physics processes such as scintillation, Cerenkov, wavelength shifting, Rayleigh scattering, and optical boundary processes with minimal approximations. Furthermore, their use can supplement or replace existing simulation techniques such as the optical libraries, and be incorporated into speeding up other parts of the simulation~\cite{Abud_2023}. This allows for a more convenient method for Monte Carlo productions after geometry changes, detector development, and testing. The use of GPUs to propagate photons is becoming more widespread, with projects such as \texttt{Chroma}~\cite{Seibert2011FastOM} being adopted in various experiments and detector simulations~\cite{Benjamin2020ChromaPR,Althueser2022, Lenardo2022nEXO127Xe}.  However, there are some limitations with using \texttt{Chroma} including no benefit from ongoing NVIDIA acceleration structure and ray tracing developments, and no use of dedicated ray tracing (RT) hardware cores in RTX GPUs. The use of the package \texttt{Opticks} resolves these mentioned issues with \texttt{Chroma} and is being incorporated into many detector simulations in particle physics, including JUNO~\cite{Opticks4} and LHCb-RICH~\cite{Li2023}. It is actively being developed and has a simple interface to \texttt{Geant4} -- with an example currently being integrated into the main repository~\cite{CatsOpticks1}.

In this work, we use the latest \texttt{Opticks} (v0.2.7) package for GPU-accelerated photon propagation for the \texttt{NEXT-CRAB-0} detector, an optical TPC with an electron-multiplying charge-coupled device (EM-CCD) camera for imaging tracks~\cite{CRAB}. We study the performance improvements with various hardware, validate sensor hits with and without the addition of \texttt{Opticks} to check its consistency with \texttt{Geant4}, and compare the images with data from the \texttt{NEXT-CRAB-0} detector.

The paper is structured as follows: in Section~\ref{sec:CRAB0}, we describe the \texttt{NEXT-CRAB-0} detector, including the latest updates to the detector. Subsequently, we detail the detector simulation in Section~\ref{sec:simulations}. Finally, in Section~\ref{sec:comparisons}, we show the performance speed improvements with different GPU architectures while also validating its performance with \texttt{Geant4}.

\section{The NEXT-CRAB-0 Detector}\label{sec:CRAB0}

\texttt{NEXT-CRAB-0} is a time projection chamber with high-pressure xenon that uses a VUV-to-visible image intensifier coupled to EM-CCD to image particle tracks. The initial demonstration of this technology along with its detailed description is given in Ref.~\cite{CRAB}.

Charged particles produced in the detector ionize and excite the xenon producing an initial flash of scintillation light (S1) along with a trail of ionization electrons along the charged particle trajectory. The ionization charge drifts in an electric field towards the electroluminescence (EL) region where the ionization charge is converted to a second flash of light in a strong electric field known as the S2 signal with gains ranging from hundreds to thousands of photons per electron~\cite{Freitas2010Scintillation}. To detect the xenon scintillation light, 174.8 $\pm$ 0.1(stat) $\pm$ 0.1(sys.)~nm~\cite{Fujii2015LiquidXenon}, the light is focused at the cathode side through a MgF$_2$ lens into an image intensifier which further amplifies to produce photons in the visible that are detected by an EM-CCD. The detector does not use wavelength shifters such as TPB or PEN commonly used in noble gas or liquid detectors which can introduce further blurring to the image~\cite{Yahlali2017,Asaadi_2019,PEN2021,Abraham_2021}. 

This work includes updates to the detector, incorporating three $^{210}$Pb needle sources inserted into the detector volume at three separate drift distances 4, 10, and 14 cm along the TPC axis, as shown in the left image of Figure~\ref{fig:Sources}. The decay chain of $^{210}$Pb leads to a $^{210}$Po daughter that alpha decays with 5.3~MeV energy. These alphas provide an excellent test of the optical simulation speeds due to the hundreds of millions of photons produced in their S1 and S2 light and are currently being studied in \texttt{NEXT-CRAB-0} for diffusion characterization in the detector. 

In addition, the single microchannel plate (MCP) \texttt{photonis} image intensifier (II) that was used in the previous study~\cite{CRAB} has been replaced with a dual MCP \texttt{photonis} II which has up to 27 times more photon gain giving brighter images. The EL region in \texttt{NEXT-CRAB-0} consists of two photo-etched hexagonal meshes with 2.5 mm diameter hexagons and 100~micron wire widths. When the optical system is in focus, hexagonal meshes can be resolved by the EM-CCD (Figure~\ref{fig:Sources}). The center image shows the image using the single MCP while the brighter image on the right shows the dual MCP II.

\section{Detector Simulation}\label{sec:simulations}

\begin{figure*}[htb]
    \centering
\includegraphics[width=0.85\textwidth]{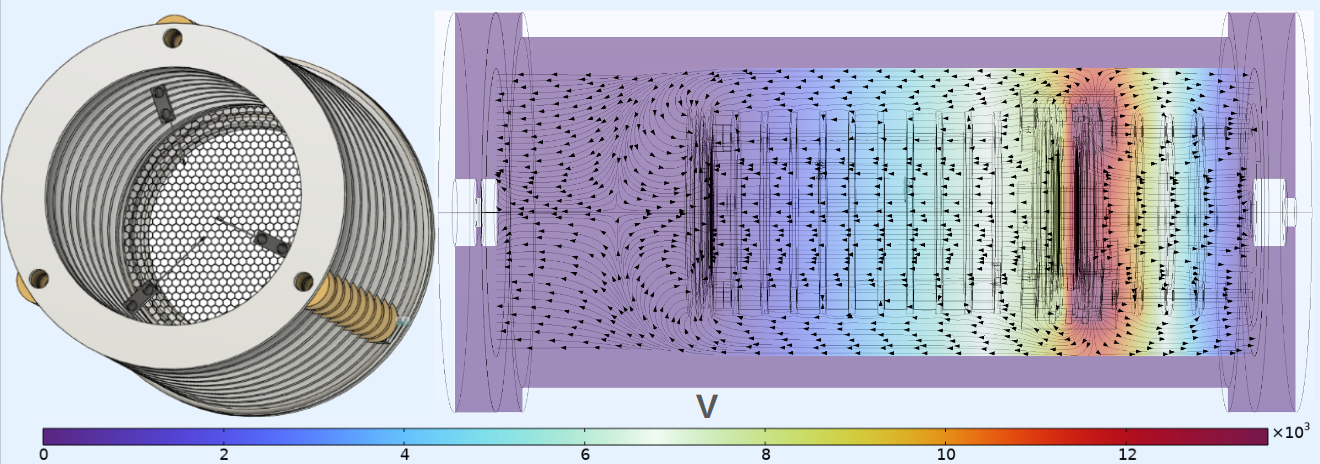}
    \caption{ Rendering of the field cage with three needle sources from \texttt{Fusion 360}, and simulated electric fields and potentials from \texttt{COMSOL}. The gradient of electric potential is shown by the color bar while the direction of the field lines is displayed by black arrows.  }
    \label{fig:COMSOL}
\end{figure*}

The detector simulation uses a modified version of \texttt{Nexus}~\cite{parmaksiz_Opticks_2023}, a \texttt{Geant4}-based simulation framework developed by the NEXT Collaboration. The modifications include a modular version of \texttt{Geant4} as described in Ref.~\cite{PFEIFFER2019121}. This enables the use of a uniform drift and electric field or employing the \texttt{Garfield++} package~\cite{garfield++} to model electron drift with an imported detailed map of the electric field generated via \texttt{COMSOL}~\cite{COMSOL}. The \texttt{COMSOL} model used here includes a CAD design (via \texttt{Fusion 360}~\cite{autodesk_fusion360}) of the \texttt{NEXT-CRAB-0} geometry including realistic hexagonal meshes and field cage. \texttt{Geant4} handles the primary particle generation with energy depositions inside the detector volume. The field cage is constructed from \texttt{Fusion 360}, which is shown in Figure~\ref{fig:COMSOL} along with the electrical fields derived from \texttt{COMSOL}. A careful description of the electric field and of the mesh is required, requiring the use of a \texttt{COMSOL} simulation to see features like the funneling of the electrons through the meshes. Alpha particles are simulated, accounting for the needle head shape which can induce geometrical effects on the alpha particle ionization cluster. The optical properties of materials used in the simulation are consistent with those in Ref.~\cite{CRAB}.

\begin{table}[h!]
    \centering
     \begin{tabular}{ |p{3.8cm}||p{2.8cm}|  }
        \hline
        Name & Version \\
        \hline
        \texttt{Geant4}  &  11.1.1 \\
        \texttt{Opticks} & 0.2.7 \\ 
        \texttt{Cuda} & 11.7 \\
        \texttt{NVIDIA OptiX } & 7.5 \\
        \texttt{NVIDIA Driver} & 535 \\
        \texttt{COMSOL} & 6.0 \\
        \texttt{Garfield++} & 4.0 \\
        \hline
    \end{tabular}
    \caption{The important packages used in this simulation along with their version numbers. \texttt{Opticks} depends on \texttt{CUDA} and \texttt{NVIDIA} GPUs. }
    \label{tab:packages}
\end{table}

The simulation can be compiled with or without \texttt{Opticks} to propagate the scintillation photons in order to compare the results with propagation via \texttt{Geant4}. The list of important packages and versions that are used in this paper are shown in Table~\ref{tab:packages}. 

\begin{figure}[h]
    \centering
\includegraphics[width=0.45\textwidth]{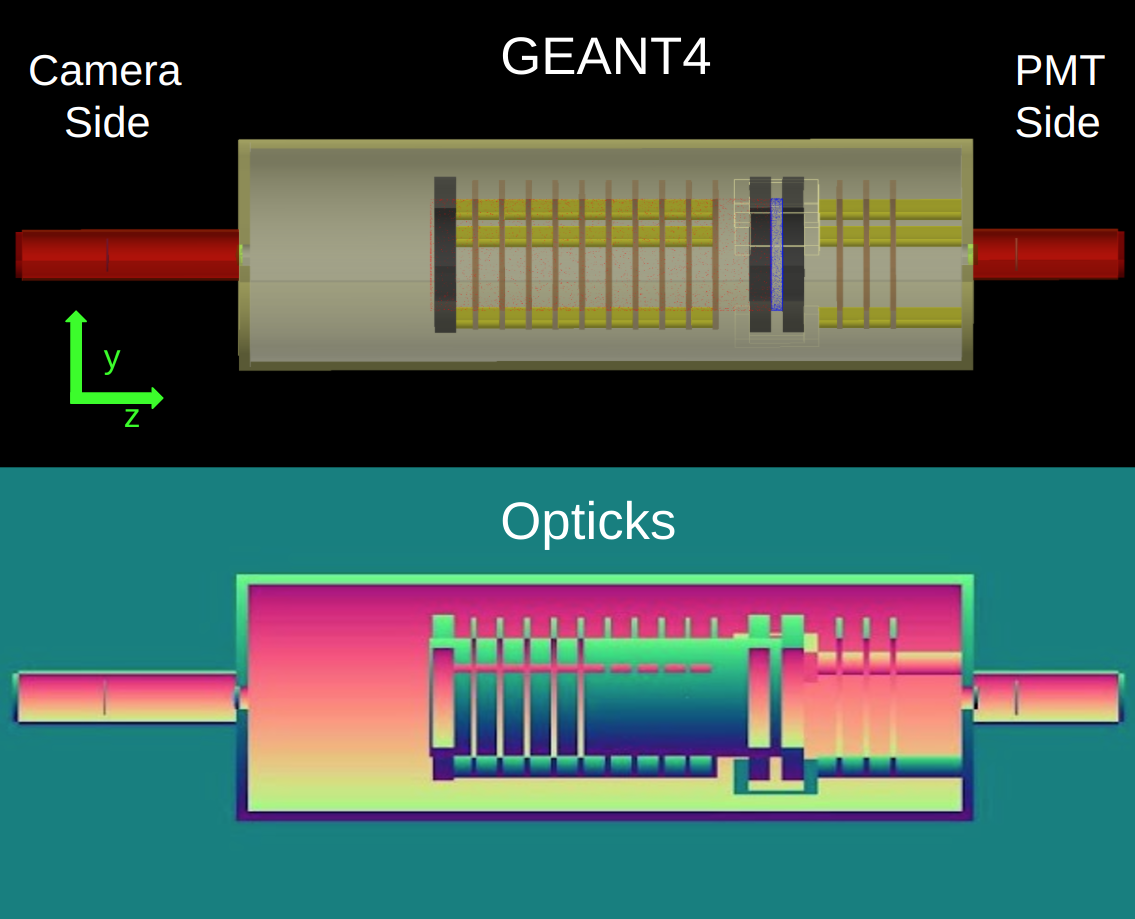}
    \caption{Geometry rendering of \texttt{NEXT-CRAB-0} in \texttt{Opticks} and \texttt{Geant4}. The $z$ direction points along the drift axis (with $z=0$ closest to the electroluminescence region), while the $x$ and $y$ coordinates are perpendicular to this. }
    \label{fig:Visual}
\end{figure}

\texttt{Opticks} is an open source project functional with the use of \texttt{NVIDIA} GPUs and has seamless integration with \texttt{Geant4} ~\cite{Opticksall,Opticks4}. It uses the ray tracing engine \texttt{OptiX 7.5} provided by \texttt{NVIDIA}~\cite{Opticks1,Opticks2,Opticks3,Opticks4}. Geometry translation and ray generation are handled by \texttt{NVIDIA OptiX} while optical physics processes such as scattering, absorption, scintillator reemission, and boundary processes are handled by \texttt{CUDA} programs~\cite{Blyth2017}. Wavelength shifting is not yet available but is currently being implemented. \texttt{Opticks} allows automatic fast geometry translation from \texttt{Geant4} to \texttt{OptiX} and with geometry conversion possible with either using a \texttt{GDML} file or \texttt{DetectorConstruction} class in \texttt{Geant4}. Comparisons of geometry translation are shown in Figure~\ref{fig:Visual}. Many geometry solids in \texttt{Geant4} such as cylinders and boxes are supported and more are currently in the process of being implemented such as tessellated solids. \texttt{Opticks} also accounts for any optical surfaces defined along with the material properties such as refractive index and absorption lengths and produces optical photons by scintillation and Cerenkov processes. In this work, we only use the scintillation process due to the negligible amount of Cerenkov light produced in the detector at particle energies of interest.

\begin{figure}[h]
    \centering
 \includegraphics[width=0.49\textwidth]{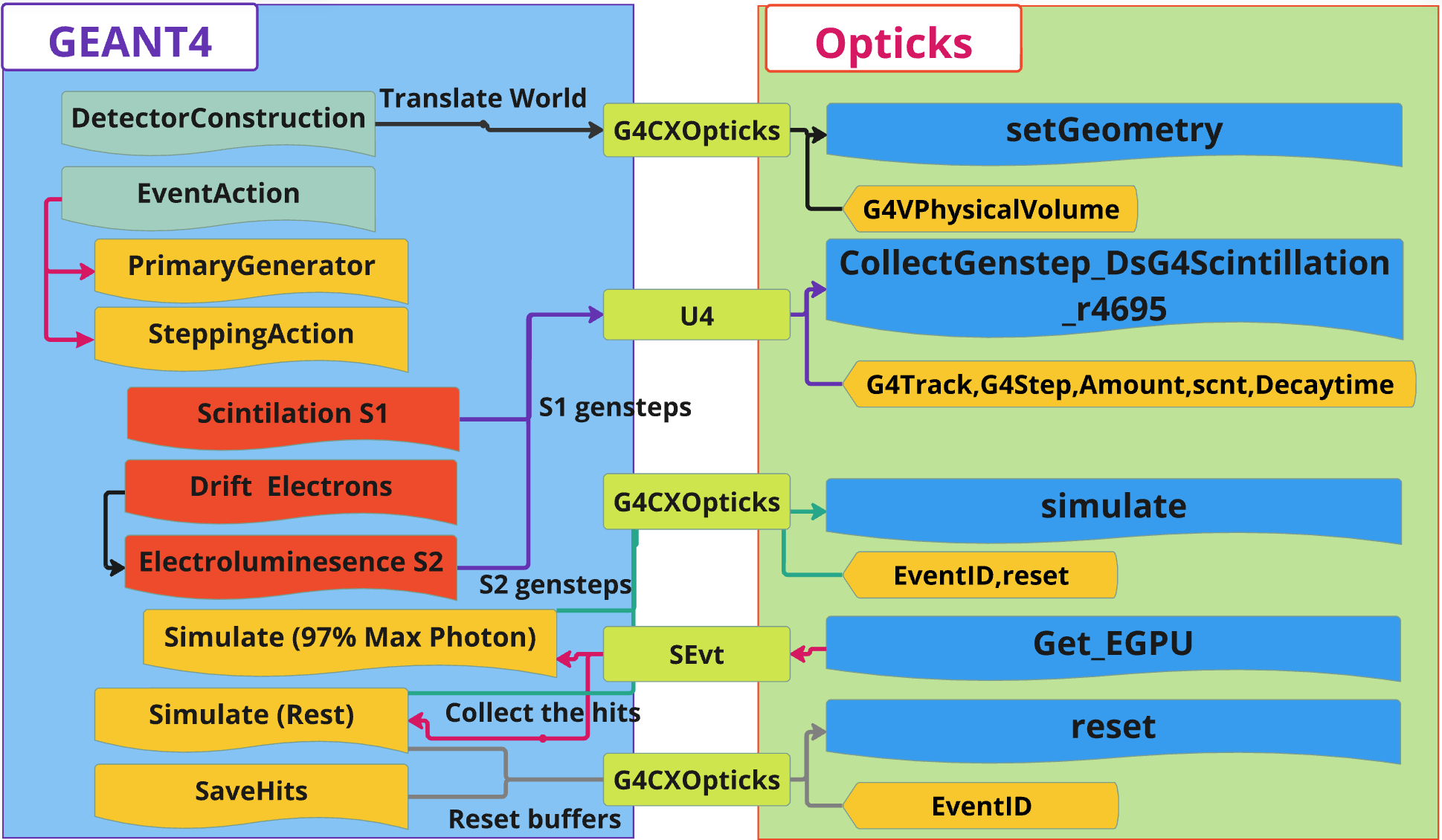}
    \caption{This flowchart depicts a \texttt{Geant4}-\texttt{Opticks} hybrid simulation pipeline. Initially, the \texttt{Geant4} geometry is translated via the \texttt{G4CXOpticks} interface. During an event, \texttt{Geant4} tracks primary and secondary particles other than optical photons. When optical photons are produced (e.g., by S1 or S2), their gensteps are transferred via the \texttt{U4} interface to \texttt{Opticks} in memory-managed batches. \texttt{Opticks} then performs GPU-accelerated simulations of these photons using its simulate function. The hits resulting from the \texttt{Opticks} simulation are collected via the \texttt{SEvt} interface component and saved. Finally, after \texttt{Geant4} processes and saves all hits, it signals \texttt{Opticks} through \texttt{G4CXOpticks} to reset for the next event.}
    \label{fig:flowchart}
\end{figure}

\texttt{Opticks} utilizes environmental variables to configure important simulation parameters, such as the maximum number of photons to simulate. 
A visual diagram of how \texttt{Opticks} interacts with our simulation package is shown in Figure~\ref{fig:flowchart}.
As an example, for an alpha event, S1 and S2 \texttt{gensteps} are collected and passed to \texttt{Opticks}. If the collected amount of photons is less than $97~\%$ of the maximum number of photons which is defined at the beginning of the simulation, photons are simulated at the end of the event. Otherwise, they are simulated in batches. The $97\%$ limit is chosen to avoid maxing out the video random access memory (VRAM) of the GPUs (listed in Table~\ref{tab:photonlimit}). After each simulation execution, the hits are collected from GPU using \texttt{SEvt::Get\_EGPU} function, and then GPU buffers are reset~\cite{simoncblyth_opticks_2024,Opticksall,Opticks4}.
If the hit collection happens at the end of the event, it is written to a file.

We assessed VRAM usage, GPU utilization, and power consumption during simulations involving 154 million photons per event. Depending on the GPU, VRAM utilization was between $80\%$ and $93\%$, contingent on the maximum photon batch capacities mentioned in Table~\ref{tab:photonlimit}, and this memory remained occupied until the simulation was complete. GPU utilization alternated between $0\%$ to $100\%$, with short inter-batch delays. High utilization corresponded to periods of active photon simulation, while the delays corresponded to transitions between batches or awaiting incoming event data. A similar pattern was observed for power consumption, with power reaching values close to the GPU's maximum during photon simulation.

\section{Simulation Comparisons}\label{sec:comparisons}
Several comparisons are performed between \texttt{Geant4} and \texttt{Opticks} where the speed with different computer architectures and simulation agreements are compared.

\subsection{Performance}\label{subsec:performance}
The performance of \texttt{Opticks} across different computer systems with varying memory and processors is investigated. We compare the single thread performance of \texttt{Geant4} photon propagation with photons propagated via a single GPU and \texttt{Opticks}. Table~\ref{tab:photonlimit} details the different computer systems used.

\begin{table}[h!]
    \centering
    \begin{tabular}{ |p{1.4cm}|p{1.5cm}|p{1.0cm}|p{1.4cm}| }
         \hline
        \makecell{GPU \\ (RTX)} & \makecell{Max \\ Photon(M)} & \makecell{VRAM \\ (GB)} & \makecell{Release \\ Date} \\
         \hline
         3050-Ti M  & 26  & 4  & 05/11/2021 \\
         2060    & 45  & 6  & 01/15/2019 \\
         3060-Ti & 60  & 8  & 12/01/2020 \\
         4090    & 160 & 24 & 12/10/2022 \\
         \hline
    \end{tabular}
    \caption{ The various GPUs studied with their available VRAM and the max number of photons batch size configured to optimize speed. }
    \label{tab:photonlimit}
\end{table}

Between $15$ and $154$ million (M) photons are simulated by generating 500 alpha particles with energy ranges of $0.5$~MeV to $5.3$~MeV at $10$~bar xenon placed in the center of the drift region. Each alpha event produces about $25,840$~photon(ph)/MeV in S1 light~\cite{Henriques2024XenonScintillation}, about $45,662$~electron(e)/MeV~\cite{Bolotnikov1997HighPressureXe} total ionization charge.  The ionization electrons are amplified in the EL region with a gain of $640$~ph/e~\cite{Freitas2010Scintillation}. The total S1 and S2 photons can reach up to 154 million for a $5.3$~MeV alpha energy. The time difference between the start and end of an alpha event was recorded excluding the loading and file writing times in this comparison. 

\begin{figure}[h]
    \centering
    \includegraphics[width=0.45\textwidth]{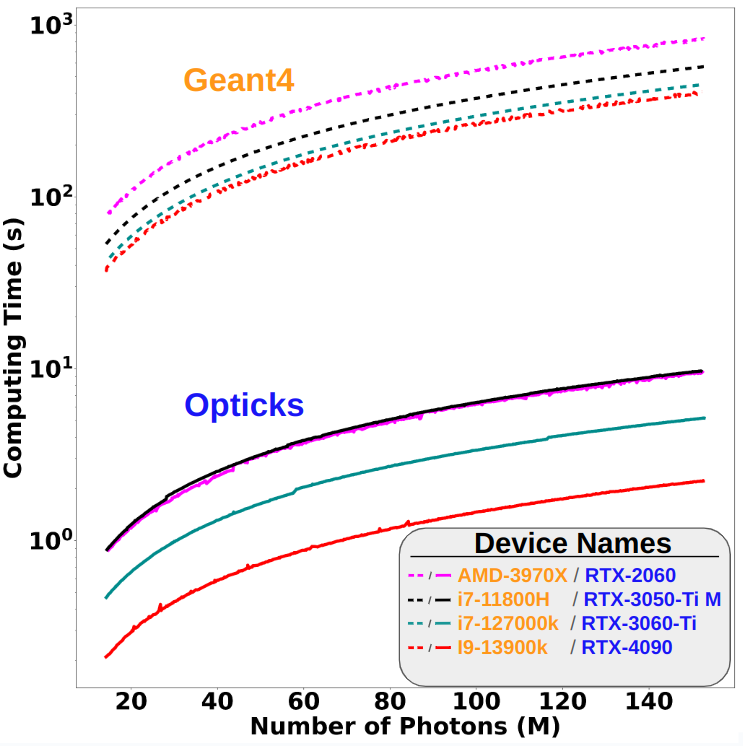}
    \caption{Performance comparison between \texttt{Geant4} and \texttt{Opticks} using different CPU (AMD/i7/i9 devices) and GPU (RTX devices) combinations. The ratios are given in Table~\ref{tab:percentages}.}
    \label{fig:Performance}
\end{figure}

The performances are shown in log-scale in Figure~\ref{fig:Performance} where it can be seen that the GPU performance is better by about a factor $58.47~\pm{0.02}$ to $181.39~\pm{0.28}$ depending on which CPU and GPU are used. These results are comparable to the recent performance result that was measured with the \texttt{CaTs} project, an \texttt{Opticks} example implementation in the main \texttt{Geant4} project~\cite{CaTsRepo,CatsOpticks1}. By comparing the slopes between the desktop GPUs RTX 20, RTX 30, and RTX 40 series, it is observed that there is an average factor of $2.09~\pm{0.36}$ speed up from one generation to the next.
Due to the power limitations of laptop GPUs and reduced clock speeds, the RTX-3050-Ti M speeds are close to the RTX-2060.
The speed improvements come from the fact that each new RTX series of GPUs is better optimized for ray tracing, has more ray tracing cores, and has higher clock speeds. 
The combination of improved GPU and CPU performance contributes to the factor of 3 speed that is seen in Table~\ref{tab:percentages}. 

The partitioning of time spent between particle generation and ionization electron drift with the optical photon propagation is also compared, excluding loading and file writing times. We simulated 10 alpha events for each device between 0.5 to 5.3 MeV energies and calculated the ratio of time spent on only generation/drift to total time (including generation, drift, and photon propagation). The mean and standard deviation of the results from CPU-only events across all devices are calculated to be $0.16\pm{0.01}\%$. Meanwhile, for events that utilize \texttt{Opticks} for photon propagation, we find this ratio to be $15.49\pm{7.24}\%$. As expected, the photon propagation dominates the simulation time, but this proportion is reduced dramatically with the introduction of \texttt{Opticks}, highlighted by the increase in fraction of time spent on the generation/drift.

\begin{table}[h!]
    \centering
     \begin{tabular}{ |p{3.5cm}||p{3.3cm}|}
         \hline
         Device Names & GPU/CPU Ratio \\
         \hline
         \texttt{RTX-3050-Ti M/i7-11800H}  &  $58.47\pm{0.02}$ \\
         \texttt{RTX-2060/AMD-3970X}  &  $86.67\pm{0.13}$ \\
         \texttt{RTX-3060-Ti/i7-12700k}  &  $86.52\pm{0.06}$ \\
         \texttt{RTX-4090/i9-13900k}  &  $181.39\pm{0.28}$ \\
         \hline
    \end{tabular}
    \caption{CPU and GPU performance ratios for each computer. The slopes are estimated by a linear fit to each curve in Figure~\ref{fig:Performance}. The ratios are calculated using the reciprocal slopes (photons/second), where the \texttt{Opticks} (GPU) result is divided by the \texttt{Geant4} (CPU) one. The uncertainties are given by the error on the line fit and propagated to the ratio.
    }
    \label{tab:percentages}
\end{table}

\subsection{Simulation Agreement}\label{subsec:hits}
The number of detected S1 and S2 photons are compared between \texttt{Opticks}, and \texttt{Geant4} at the camera and PMT in \texttt{NEXT-CRAB-0} as labeled in Figure~\ref{fig:Visual}. Both sensors are modeled as circular disks, and we use a uniform electric field in these comparisons as described in Ref.~\cite{CRAB}. We set the efficiency of the photodetectors to 1 and simulated 1000 alpha particles with $5.3$~MeV energy on the side of the field cage. Comparisons are done using the \texttt{i9-13900k} CPU and \texttt{RTX-4090} GPU. 

\begin{figure}[h]
    \centering
    \includegraphics[width=0.39\textwidth]{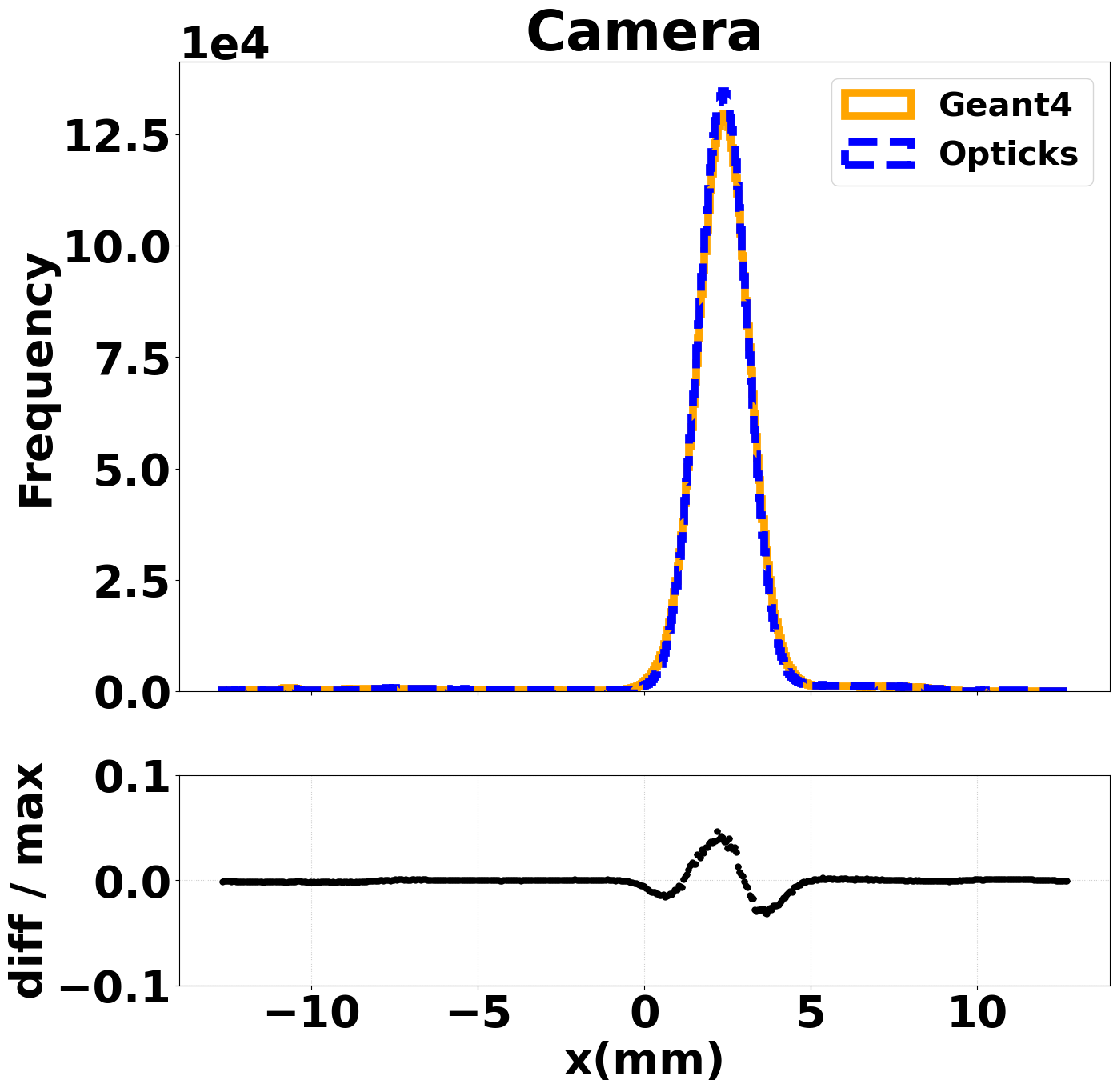}
    \includegraphics[width=0.39\textwidth]{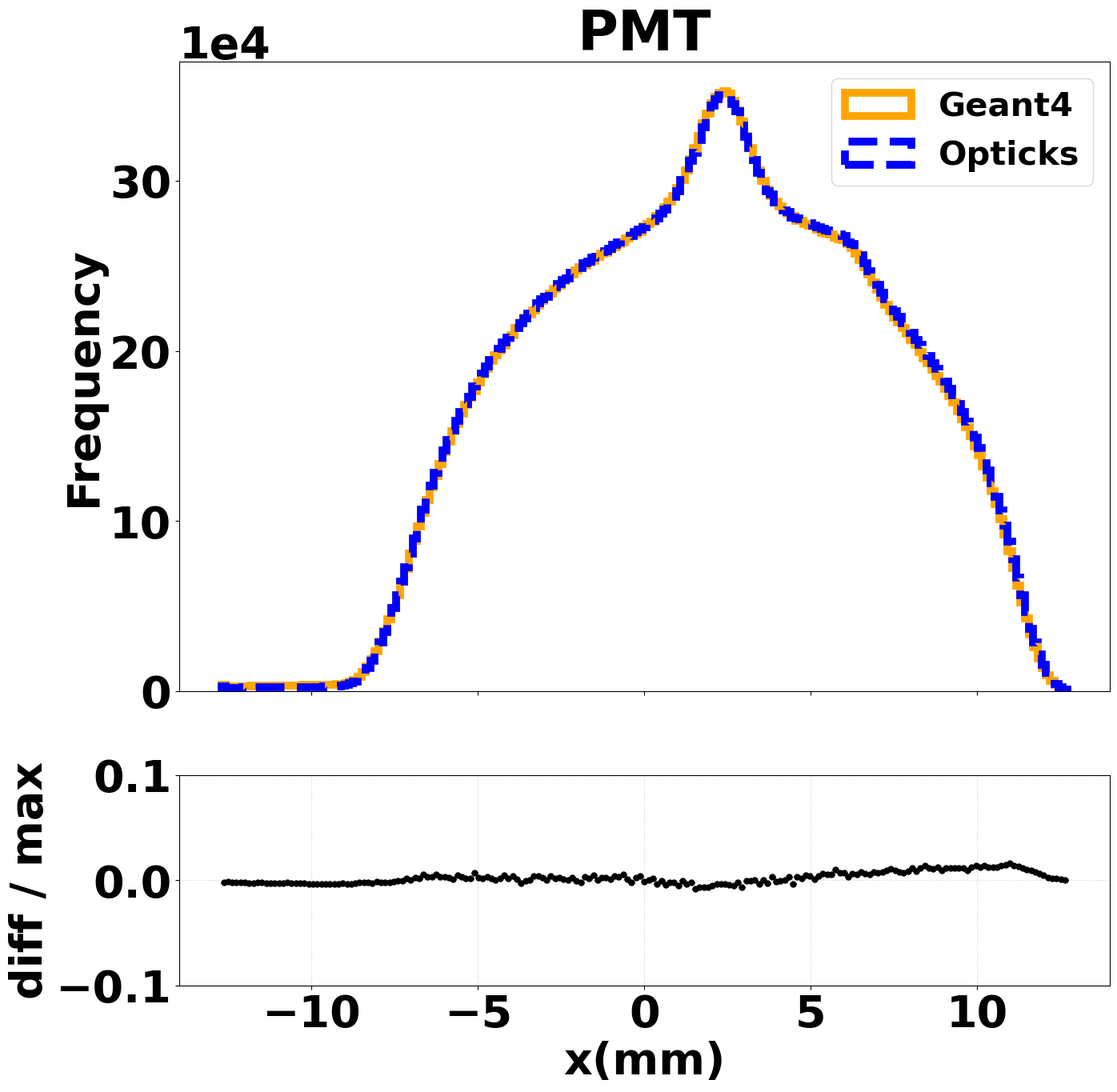}
    \caption{Distribution in $x$ of detected S2 photons for the camera (top) and PMT (bottom) planes, simulated using \texttt{Geant4} and \texttt{Opticks}. Both simulations are in good agreement with one another. }
    \label{fig:S2_hits}
\end{figure}

Figure~\ref{fig:S2_hits} shows the $x$ coordinate of detected S2 photons at the camera and PMT planes, histogrammed over all events for \texttt{Geant4} and \texttt{Opticks}. Good agreement is seen in the PMT and camera plane, with similar agreement observed in the distributions of the $y$ coordinate.

We also compared S1 and S2 photon hits at the camera and PMT. The distributions of photon numbers over the 1000 alpha events are fit to a Gaussian distribution, and their means and standard deviations are reported in Table~\ref{tab:reflections}. We find that the number of detected S1 and S2 photons is in agreement for \texttt{Geant4} and \texttt{Opticks}, within the statistical uncertainty.

It should be noted that in the geometry construction, we found that boundary overlaps in the \texttt{Geant4} geometry can lead to discrepancies in the propagation of optical photons between \texttt{Geant4} and \texttt{Opticks}. In addition, touching volumes with different mother volumes can also lead to different inheritance of optical properties, generating a difference in the simulations.

\begin{table}[h!]
    \centering
    \makebox[0.45\textwidth][c]{ 
        \begin{tabular}{|c|c|c||c|c|}
            \hline
            & \multicolumn{2}{c||}{\textbf{PMT}} & \multicolumn{2}{c|}{\textbf{Camera}} \\ 
            \hline
            & \textbf{S1} & \textbf{S2} & \textbf{S1} & \textbf{S2} \\ 
            \hline
            \textbf{Geant4} & 66$\pm$9 & 168957$\pm$880 & 40$\pm$7 & 25477$\pm$157 \\ 
            \textbf{Opticks} & 65$\pm$9 & 169745$\pm$1056 & 41$\pm$6 & 25528$\pm$227 \\ 
            \hline
        \end{tabular}
    }
    
 
    
    \caption{The mean number of photons detected at the PMT and camera for S1 and S2 photons. The uncertainties are the one-sigma values from Gaussian fits of the photon distribution. }
    \label{tab:reflections}
\end{table}

\begin{figure}[tb]
    \centering
    \includegraphics[width=0.45\textwidth]{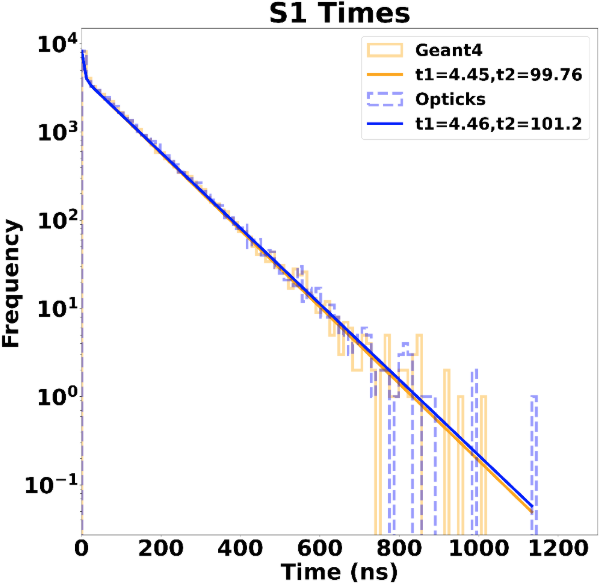}
    \includegraphics[width=0.45\textwidth]{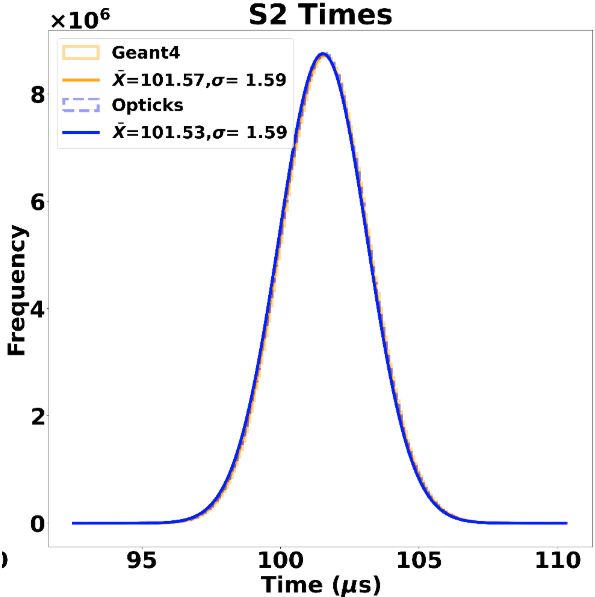}

    \caption{Distributions of arrival times for S1 (top) and S2 (bottom) photons. The $t1$ and $t2$ correspond to the fitted fast and slow scintillation times, respectively. The $\bar{X}$ and $\sigma$ indicate the mean and standard deviation of the S2 photon arrival times. Both simulations are in good agreement with one another.   }
    \label{fig:times}
\end{figure}

Comparisons of the wavelength and the timing constants of the S1 and S2 light were also done. We found that the wavelength agreement was within $0.1\%$. The distribution of the S1 photon arrival times was fitted with a sum of two exponential decays, accounting for the fast and slow scintillation time constants in gaseous xenon. This was found to be within 3\% for the fast component and 1.5\% for the slow component. The mean S2 arrival times are in agreement to within a fraction of a percent (see Figure~\ref{fig:times}).

\begin{figure}[tb]
    \centering
    \includegraphics[width=0.48\textwidth]{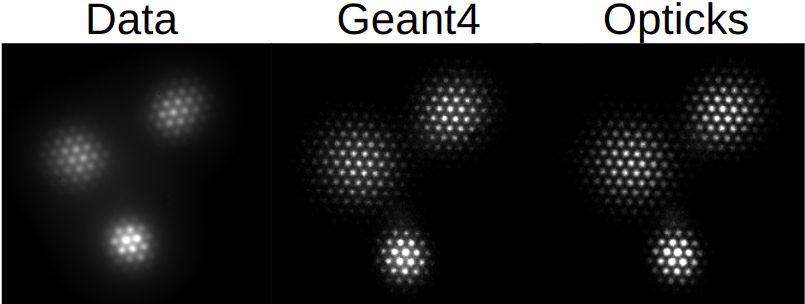}
    \caption{Comparisons of the average image of 5.3 MeV alpha particles recorded by the \texttt{NEXT-CRAB-0 detector} with \texttt{Geant4} and \texttt{Opticks}. }
    \label{fig:VisualData}
\end{figure}

Finally, a comparison of the images produced by \texttt{Geant4} and \texttt{Opticks} simulations at the Camera is shown in Figure~\ref{fig:VisualData}. For a visual comparison, we also include an image of 5.3 MeV alpha particles recorded at 8-bar xenon in \texttt{NEXT-CRAB-0}. The hexagonal shapes in the images are obtained due to the funneling of the ionization electrons through the hexagonal mesh. This feature is obtained through the detailed model of the electric field lines and electron drift via \texttt{Garfield++} and \texttt{COMSOL}. The oblong shape of the upper right alpha image is due to distortion from the needle, and the simulation has a slightly different mesh rotation compared with the data. The benefit of the speed improvements enabled by this work has allowed for more fine-tuned models, such as using more detailed electrical field maps, to be studied more easily.


\section{Summary }
 In summary, we have compared the performance of \texttt{Opticks}, an \texttt{NVIDIA OptiX API} 7.5 GPU-accelerated photon propagation compared with a single-threaded \texttt{Geant4} simulation. From these results, we find that \texttt{Opticks} improves simulation speeds ranging from $58.47~\pm{0.02}$ to $181.39~\pm{0.28}$ over \texttt{Geant4} depending on the computational specifications.
 Performance improves with the generation of GPUs and from one generation to the next, an average of $2.09\pm{0.35}$ factor speed improvements is observed. Comparison of S1 and S2 response of \texttt{Opticks} to \texttt{Geant4} shows that detection wavelengths and times agree. We also find that the total photon hits are in good agreement.
 
\bmhead{Acknowledgments}
The authors extend their sincere appreciation to Simon Blyth for his assistance regarding the \texttt{Opticks} package.
The NEXT Collaboration acknowledges support from the following agencies and institutions: the European Research Council (ERC) under Grant Agreement No. 951281-BOLD; the European Union’s Framework Programme for Research and Innovation Horizon 2020 (2014–2020) under Grant Agreement No. 957202-HIDDEN; the MCIN/AEI of Spain and ERDF A way of making Europe under grants PID2021-125475NB and RTI2018-095979, and the Severo Ochoa and Mar\'ia de Maeztu Program grants CEX2023-001292-S, CEX2023-001318-M and CEX2018-000867-S; the Generalitat Valenciana of Spain under grants PROMETEO/2021/087 and CISEJI/2023/27; the Department of Education of the Basque Government of Spain under the predoctoral training program non-doctoral research personnel; the Spanish la Caixa Foundation (ID 100010434) under fellowship code LCF/BQ/PI22/11910019; the Portuguese FCT under project UID/FIS/04559/2020 to fund the activities of LIBPhys-UC; the Israel Science Foundation (ISF) under grant 1223/21; the Pazy Foundation (Israel) under grants 310/22, 315/19 and 465; the US Department of Energy under contracts number DE-AC02-06CH11357 (Argonne National Laboratory), DE-AC02-07CH11359 (Fermi National Accelerator Laboratory), DE-FG02-13ER42020 (Texas A\&M), DE-SC0019054,  DE-SC0019223 and DE-SC0024438 (Texas Arlington); the US National Science Foundation under award number NSF CHE 2004111; the Robert A Welch Foundation under award number Y-2031-20200401. Finally, we are grateful to the Laboratorio Subterr\'aneo de Canfranc for hosting and supporting the NEXT experiment.

\bibliography{biblio}

\end{document}